\documentstyle[aps,preprint]{revtex}

\def\[{\left\lbrack}
\def\]{\right\rbrack}

\def\({\left(}
\def\){\right)}
\def\ih{\'\i}

\begin{document}

\title{Removing the Wess Zumino fields in the BFFT formalism}

\author{Jorge Ananias Neto\thanks{jorge@fisica.ufjf.br}}
\address{Departamento de F\ih sica, ICE,\\ Universidade Federal de Juiz de Fora, 36036-900, Juiz de Fora, MG, Brazil }

\maketitle

\begin{abstract}
In this paper we give some prescriptions in order to remove the Wess Zumino fields of the BFFT formalism and, consequently, we derive a gauge invariant system written only in terms of the original second class phase space variables. Here, the Wess Zumino fields  are considered only as auxiliary variables that permit us to reveal the underlying symmetries present in a second class system. We apply our formalism in three important and illustrative constrained systems which are the Chern Simons Proca theory, the Abelian Proca model and the reduced SU(2) Skyrme model.
\end{abstract}
\vskip 1 cm
PACS numbers: 11.10.Ef; $\,$11.15.-q $\\$

\setlength{\baselineskip}{20pt}  
\renewcommand{\baselinestretch}{2}
\newpage

\section{Introduction}

The BFFT formalism\cite{BFFT,embed} converts second class constrained systems into first class ones by enlarging the original second class phase space variables with the Wess Zumino (WZ) fields. In order to guarantee that the same degrees of freedom are maintained with the original second class system, the WZ fields are introduced in equal number to the number of second class constraints. The introduction of the WZ fields modifies the second class constraints and the second class Hamiltonian in order to satisfy a first class algebra. Thus, the presence of the WZ fields allows us to obtain a gauge invariant model where symmetries are revealed from the original second class system. The symmetries permit us to describe the physical properties in a more general way. For this reason we can disclose important and interesting physical results. As an example, we can cite the case of a noncommuting second class algebra resulting from a nonstandard gauge condition\cite{rep,rep2}.

The purpose of this paper is to give a prescription in order to remove the WZ fields of the BFFT formalism and, consequently, to derive a gauge invariant system written only in terms of the original second class phase space variables. Our final results are similar to those obtained by employing the gauge unfixing formalism (GU formalism)\cite{MR,Vyt} in which has the property that does not extend the phase space in converting the second class systems into first class ones. In our method, the WZ fields are treated as auxiliary variables that permit us to build a first class system from the second class one, and, consequently, to enforce symmetries. As an additional step, we replace the WZ fields by convenient functions that lead us to derive a first class system written only in terms of the initial second class phase space variables. As we will see, we can choose gauge symmetry generators and, consequently, gauge fixing conditions that allow us to reveal interesting physical properties. Since many important constrained systems have only two second class constraints, so, in this paper, we describe our formalism only for systems with two second class constraints without any loss of generality. 

In order to clarify the exposition of the subject, this paper is organized as follows: In Section II we give a short review of the BFFT formalism. In Section III, we present the formalism. In Section IV, we apply the formalism to the Chern Simons Proca theory (CSP)\cite{CSP}, the Abelian Proca model\cite{Vyt2} and the collective coordinates expansion of the SU(2) Skyrme model\cite{Skyrme,ANW}. These three physical systems are important nontrivial examples of the second class constrained systems. The Chern Simons Proca theory concerns with the interaction of a charged particle with magnetic field and it is known that this model exhibits a noncommutative algebra\cite{NCG}. 
The Abelian Proca model is a four dimensional field theory which describes electromagnetism with massive photon field. The Skyrme model is a nonlinear effective field theory which describes hadrons physics and its quantization is obtained with quantum mechanics on a curved space. Here, we would like to remark that, using our formalism, we have obtained a noncommutative Skyrmions system, a new result which is derived from a particular gauge condition. In Section V, we make our concluding remarks.

\section{A brief review of the BFFT formalism}

As we have mentioned in the introduction, the BFFT formalism converts second class system into first class one by adding WZ fields to the original second class system. All the second class constraints and the second class Hamiltonian are changed in order to satisfy a first class algebra.

Consider the original phase space variables as $(q_i,p_i)$ where a constrained system has two second class constraints, $T_\alpha, \,\,\alpha=1,2$, obeying the algebra

\begin{equation}
\{T_\alpha,T_\beta\}=\Delta_{\alpha\beta},
\end{equation}
where the matrix $\,\Delta_{\alpha\beta}\,$ has a nonvanishing determinant. 
First, in the BFFT formalism, the two first class constraints are constructed by the following expansion

\begin{eqnarray}
\label{exp1}
\tilde{T}_\alpha(q_i,p_i,\Phi_\alpha)=T_\alpha+\sum_{m=1}^\infty  T_\alpha^{(m)},
\end{eqnarray}
where $\Phi_\alpha$ are the WZ fields satisfying the algebra

\begin{equation}
\label{wb}
\{\Phi_\alpha,\Phi_\beta\}=\omega_{\alpha\beta},
\end{equation}
being $\omega_{\alpha\beta}$ an antisymmetric matrix. 
$T_\alpha^{(m)}$ are the correction terms which are powers of $\Phi_\alpha\, ,i.e.,\, T_\alpha^{(m)}\sim \Phi_\alpha^{(m)}$. The first class constraints must satisfy the boundary condition
\begin{equation}
\tilde{T}_\alpha(q_i,p_i,0)=T_\alpha^{(0)}=T_\alpha.
\end{equation}
From the Abelian first class algebra

\begin{equation}
\{\tilde{T}_\alpha,\tilde{T}_\beta\}=0,
\end{equation}
we obtain recursive equations which determine the correction terms $T_\alpha^{(m)}$. As an example, we have a basic equation in the lowest order 

\begin{equation}
\label{basic}
\Delta_{\alpha\beta}+X_{\alpha\gamma}\,\omega^{\gamma\lambda}\,X_{\lambda\beta}=0,
\end{equation}
and the first order correction term written as

\begin{equation}
T_\alpha^{(1)}=X_{\alpha\beta}(q_i,p_i)\Phi^\beta.
\end{equation}
The matrices $\omega_{\alpha\beta}$ and $X_{\alpha\beta}$ in Eqs.(\ref{wb}) and (\ref{basic}), which are the inherent arbitrariness of the BFFT formalism, can be chosen with the aim of obtaining algebraic simplifications in the determination of the correction terms $T_\alpha^{(m)}$.

In a similar way, the gauge invariant Hamiltonian is obtained by the expansion 

\begin{equation}
\label{bfh}
\tilde{H}=H_c+\sum_{m=1}^\infty H^{(m)},
\end{equation}
where $H_c$ is the canonical second class Hamiltonian and the correction terms, $H^{(m)}$, are powers of $\Phi_\alpha\, ,i.e.,\, H^{(m)}\sim \Phi_\alpha^{(m)}$. Also, from the Abelian first class algebra

\begin{equation}
\{\tilde{H}, \tilde{T}_\alpha\}=0,
\end{equation}
we have recursive equations which determine the correction terms $H^{(m)}$ and, consequently, the gauge invariant Hamiltonian. 

\section{Removing the Wess Zumino fields}

Our formalism begins by choosing, as example, $\tilde{T}_1$, one of the two first class constraints, Eq.(\ref{exp1}), to be the extended gauge symmetry generator of the theory 

\begin{equation}
\label{egsg}
\tilde{T}=\tilde{T}_1.
\end{equation}
The other first class constraint, $\tilde{T}_2$, will be discarded. To eliminate the WZ auxiliary fields, $\,\Phi_\alpha$, we must find functions for the WZ fields written only in terms of the original second class phase space variables $(q_i,p_i) $, namely 

\begin{equation}
\label{fct}
\Phi_\alpha=F_\alpha(q_i,p_i).
\end{equation} 
At this stage, two conditions must be satisfied: the first one determines that the algebraic form of the functions $F_\alpha(q_i,p_i)$ must have the same infinitesimal gauge transformations given by  $\Phi_\alpha$, i.e.

\begin{equation}
\label{rep}
\delta\Phi_\alpha=\delta F_\alpha(q_i,p_i),
\end{equation}
where

\begin{equation}
\delta\Phi_\alpha=\epsilon \{ \Phi_\alpha, {\tilde T} \},
\end{equation}
and

\begin{equation}
\label{dfa}
\delta F_\alpha=\epsilon \{ F_\alpha, T_1 \},
\end{equation}
being $\,\epsilon\,$ an infinitesimal parameter and 
$T_1$ the second class constraint that builds the extended gauge symmetry generator; the second condition imposes that when we make the constraint surface $ T_2=0$, where $T_2$ is the original second class constraint that builds the discarded first class constraint, the function $F_\alpha(q_i,p_i)$ must vanish, i.e.

\begin{eqnarray}
\label{bound}
T_2=0 \; \Rightarrow \; F_\alpha(q_i,p_i)=0.
\end{eqnarray}
With this condition we must recover the second class Hamiltonian, $H_c$. The relation (\ref{bound}) is the boundary condition of the formalism or the gauge fixing constraint that reduces our gauge invariant model to the second class one. This condition ensures the equivalence of the gauge invariant model obtained by our prescription and the original second class theory that has been embedded by the BFFT formalism\cite{MR}.

It is important to mention that we have arbitrariness in our prescription because we need to select one of the two first class constraints, Eqs.(\ref{exp1}), to be the extended gauge symmetry generator. In addition, the two conditions exposed above, at first, do not determine completely the algebraic form of the function $F_\alpha(q_i,p_i)$. However, arbitrariness, in principle, occurs in all methods that embed second class constrained systems
and can be useful to unveil important physical properties of the models.  

Substituting Eq.(\ref{fct}) in the BFFT first class Hamiltonian, Eq.(\ref{bfh}), we obtain a gauge invariant Hamiltonian, $\tilde H$, written only as a function of the original second class phase space variables $ (q_i,p_i) $, satisfying the first class algebra

\begin{eqnarray}
\label{fch2}
\{ \tilde{H}, T_1 \} =0,\\
\label{fcc}
\{ T_1, T_1 \} =0,
\end{eqnarray}
where now the second class $T_1$ becomes the only gauge symmetry generator of the theory. The relations (\ref{fch2}) and (\ref{fcc}) show that we have achieved the same results of the GU first class algebra. Thus, we can conclude that our formalism also connects the BFFT method with the GU formalism.

\section{Applications of the formalism}

\subsection{The Chern Simons Proca theory}

The Chern Simons Proca theory (CSP) describes a charged particle constrained to move on a two dimensional plane, interacting with a constant magnetic field $B$ which is orthogonal  to the plane. In the vanishing mass limit (infrared limit), the Lagrangian that governs the dynamics is

\begin{equation}
L={B\over 2} q_i\epsilon_{ij}\dot{q_j}-{k\over 2}q_iq_i,
\end{equation}
where $k$ is a constant and $\epsilon_{12}=1$. The CSP model is a second class constrained system with the two constraints given by

\begin{equation}
T_i=p_i+{B\over 2}\epsilon_{ij} q_j, \;\; i=1,2
\end{equation}
where $p_i$ are the canonical momenta ($p_i={\partial L\over \partial \dot{q}_i}$), and the Poisson brackets between the second class constraints read as

\begin{equation}
\{T_i,T_j\}=\Delta_{ij}=B\epsilon_{ij}.
\end{equation}
From the Legendre transformation we obtain the second class Hamiltonian

\begin{equation}
\label{hcsp}
H_c=p_i\dot{q}_i-L={k\over 2} q_iq_i.
\end{equation}

Using the BFFT formalism to convert this second class system into first class one, we get the two first class constraints and the gauge invariant Hamiltonian written as\cite{PKPKK}

\begin{eqnarray}
\label{f1csp}
\tilde{T}_1&=&T_1+\sqrt{B}\,c_1,\\
\label{f2csp}
\tilde{T}_2&=&T_2+\sqrt{B}\,c_2,\\
\label{fhcsp}
\tilde{H}&=&{k\over 2}[ q_iq_i+{2\over\sqrt{B}}\epsilon_{ij}c_iq_j
+{1\over B}c_ic_i ],
\end{eqnarray}
where $\,c_1$ and $\,c_2$ are the WZ variables. By construction, we have a first class algebra

\begin{eqnarray}
\{\tilde{T_i},\tilde{T_j}\}&=&0,\\
\{\tilde{H},\tilde{T}_i\}&=0&,
\end{eqnarray}
where the WZ variables satisfy the following Poisson brackets

\begin{equation}
\{c_i,c_j\}=\epsilon_{ji}.
\end{equation}

At this point, we begin our formalism by choosing the first class constraint, Eqs.(\ref{f1csp}), to be the extended gauge symmetry generator

\begin{equation}
\label{esgcsp}
\tilde{T}=\tilde{T}_1=T_1+\sqrt{B}\,c_1=p_1+{B\over 2}q_2+\sqrt{B}\,c_1.
\end{equation}
The infinitesimal gauge transformations of the WZ variables generated by the extended gauge symmetry generator $\tilde{T}$ are

\begin{eqnarray}
\label{dc1}
\delta c_1&=&\epsilon\{c_1,\tilde{T}\}=\epsilon\{c_1,p_1+{B\over 2}q_2+\sqrt{B}\,c_1\}=0,\\
\label{dc2}
\delta c_2&=&\epsilon\{c_2,\tilde{T}\}=\epsilon\{c_2,p_1+{B\over 2}q_2+\sqrt{B}\,c_1\}=\epsilon\;\sqrt{B}.
\end{eqnarray}
From the infinitesimal gauge transformations, Eq.(\ref{dc1}), we can choose a representation for $c_1$ as

\begin{equation}
\label{rc1}
c_1=0.
\end{equation}
A representation for $c_2$ can be determined by imposing the first class strong equation, Eq.(\ref{f2csp})

\begin{eqnarray}
\label{rc2}
T_2+\sqrt{B}\, c_2=0 \;\Rightarrow \;c_2=-{1\over \sqrt{B}}T_2.
\end{eqnarray}
As we can see, the function for $c_2$ satisfies the infinitesimal gauge transformation, Eq.(\ref{dc2}),

\begin{eqnarray}
\delta c_2=\epsilon\{c_2,\tilde{T}\}=\epsilon\{-{1\over \sqrt{B}}T_2,\tilde{T}\}=\epsilon\{-{1\over \sqrt{B}}T_2,T_1\}=\epsilon\;\sqrt{B}.
\end{eqnarray}
Then, substituting the functions for $c_1$ and $c_2$, Eqs.(\ref{rc1}) and (\ref{rc2}), in the first class Hamiltonian, Eq.(\ref{fhcsp}), we obtain a gauge invariant Hamiltonian written only in terms of the original second class phase space variables

\begin{eqnarray}
\label{fhcsp2}
\tilde{H}={k\over 2}q_iq_i+{k\over B}q_1T_2+{k\over 
2B^2}T_2^2={k\over 2}\[q_2q_2+\(q_1+{T_2\over B}\)^2\],
\end{eqnarray}
being the only gauge symmetry generator 

\begin{equation}
T_1=p_1+{B\over 2}q_2,
\end{equation}
which satisfies an Abelian first class algebra

\begin{eqnarray}
\{ T_1, T_1 \} =0,\\
\{\tilde{H},T_1\}=0.
\end{eqnarray}
We can observe that when we make $\,T_2=p_2-{B\over 2}q_1=0$ (the second class constraint that builds the discarded first class constraint, condition two of the formalism) the gauge invariant Hamiltonian, Eq.(\ref{fhcsp2}), reduces to the CSP second class Hamiltonian, Eq.(\ref{hcsp}). Moreover, the invariant Hamiltonian, Eq.(\ref{fhcsp2}), is the same obtained when we use another gauge invariant formalism\cite{GS}.

\subsection{The Abelian Proca model}
The Abelian Proca model is a four dimensional field theory with the corresponding Lagrangian density given by

\begin{equation}
{\cal L}= -{1\over 4}F_{\mu\nu}F^{\mu\nu}
+ {m^2\over 2} A^\mu A_\mu,
\end{equation}
where $g_{\mu\nu}=diag(+,-,-,-)$ and $F_{\mu\nu}=\partial_\mu A_\nu-\partial_\nu A_\mu$. The explicit mass term breaks the gauge invariance and, consequently, we have a second class constrained system. The primary constraint is

\begin{equation}
\label{prip}
T_1=\pi_0\approx 0.
\end{equation}
By using the Legendre transformation we obtain the canonical Hamiltonian written as

\begin{equation}
\label{hcp}
H_c=\int d^3x {\cal H}_c=\int d^3x \;[ {1\over 2} \pi_i\pi_i + {1\over 4}F_{ij}F_{ij}-{m^2\over 2}(A_0^2-A_i^2)+A_0(\partial_i\pi_i)],
\end{equation}
with $\pi_i={\partial L\over \partial \dot{A}_i}=-F_{0i}$. From the temporal stability condition of the primary constraint, Eq.(\ref{prip}), we get the secondary constraint 

\begin{equation}
T_2=-\partial_i\pi_i+m^2 A_0\approx 0.
\end{equation}
We observe that no further constraints are generated via this iterative procedure. Then, $T_1$ and $T_2$ are the total second class constraints of the Abelian Proca model.

Using the BFFT formalism to convert this second class system into first class one, we obtain the two first class constraints and the gauge invariant Hamiltonian written as\cite{Vyt2}

\begin{eqnarray}
\label{f1p}
\tilde{T}_1&=&T_1+m^2\theta,\\
\label{f2p}
\tilde{T}_2&=&T_2+\pi_\theta,\\
\label{fhp}
\tilde{H}&=&H_c+\int d^3 x \,[ {\pi_\theta^2\over 2m^2}+{m^2\over2} (\partial_i\theta)^2-m^2\theta\,\partial_i A_i ],
\end{eqnarray}
where the extra canonical pair of fields $\theta$ and $\pi_\theta$ satisfy the algebra $\{\theta(x),\pi_\theta(y)\}=\epsilon \delta(x-y)$. The first class constraints and the first class Hamiltonian obey the following Poisson brackets

\begin{eqnarray}
\{\tilde{T}_1,\tilde{T}_2\}&=&0,\\
\{\tilde{T}_1,\tilde{H}\}&=&\tilde{T}_2,\\
\{\tilde{T}_2,\tilde{H}\}&=&0.
\end{eqnarray}

In order to apply our formalism,  we choose the first class constraint, Eq.(\ref{f2p}), to be the extended gauge symmetry generator 

\begin{equation}
\tilde{T}=\tilde{T}_2= T_2+\pi_\theta=-\partial_i\pi_i+m^2 A_0+\pi_\theta.
\end{equation}
The infinitesimal gauge transformations of the WZ fields generated by the extended gauge symmetry generator $\tilde{T}$ are
\begin{eqnarray}
\label{dtheta}
\delta\theta&=&\epsilon\{\theta,-\partial_i\pi_i+m^2 A_0+\pi_\theta\}=\epsilon\{\theta(x),\pi_\theta(y)\}=\epsilon\,\delta(x-y),\\
\label{dpitheta}
\delta\pi_\theta&=&\epsilon\{\pi_\theta,-\partial_i\pi_i+m^2 A_0+\pi_\theta\}=\epsilon\{\pi_\theta(x),\pi_\theta(y)\}=0.
\end{eqnarray}
From the infinitesimal gauge transformations, Eq.(\ref{dpitheta}), we can choose a representation for $\pi_\theta$ as

\begin{equation}
\label{pitheta}
\pi_\theta=0.
\end{equation}
A representation for $\theta$ can be determined by imposing the first class strong equation, Eq.(\ref{f1p}),

\begin{eqnarray}
\label{theta}
 T_1+m^2\theta=0 \Rightarrow\; \theta=-{1\over m^2}T_1=-{1\over m^2}\pi_0.
\end{eqnarray}
As we can observe, the function for $\theta$ satisfies the infinitesimal gauge transformation, Eq.(\ref{dtheta}),

\begin{eqnarray}
\delta\theta=\epsilon\{\theta,\tilde{T}\}=\epsilon\{-{1\over m^2}\pi_0,T_2\}=\epsilon\{-{1\over m^2}\pi_0,-\partial_i\pi_i+m^2 A_0\}=\epsilon\delta(x-y).
\end{eqnarray}
Substituting the WZ formulas, Eqs.(\ref{pitheta}) and (\ref{theta}), in the extended first class Hamiltonian, Eq.(\ref{fhp}), we get a first class Hamiltonian written only in terms of the original second class fields
  
\begin{equation}
\label{fhp1}
\tilde{H}=H_c+\int d^3x\, \[ \pi_0\,\partial_iA_i + {1\over 2 m^2}(\partial_i\pi_0)^2\],
\end{equation}
or

\begin{equation}
\label{fhp2}
\tilde{H}=H_c+\int d^3x\, \[ \pi_0\,\partial_iA_i - {1\over 2 m^2}\pi_0\,\partial^2_i\pi_0\],
\end{equation}
being the only gauge symmetry generator  

\begin{equation}
T_2=-\partial_i\pi_i+m^2 A_0,
\end{equation}
which satisfies an Abelian first class algebra

\begin{eqnarray}
\{ T_2, T_2 \} =0,\\
\{\tilde{H},T_2\}=0.
\end{eqnarray}
If we make the second class constraint equal to zero, $\,\pi_0=0\,$, then we can observe that the first class Hamiltonian, Eq.(\ref{fhp2}), reduces to the original second class Hamiltonian, Eq.(\ref{hcp}). In addition, the first class Hamiltonian, Eq.(\ref{fhp2}), is identical to the gauge invariant Hamiltonian which was derived by using the GU formalism\cite{Vyt2}.

\subsection{The reduced Skyrme model or the Skyrme model expanded in terms of the SU(2) collective coordinates}

The Skyrme model describes baryons and their interactions through soliton solutions of the nonlinear sigma model type Lagrangian given by

\begin{eqnarray}
\label{Sky}
L = \int \, d^3x \[ {f_\pi^2\over 4} Tr\, (\partial_\mu U \partial^u U^+) + {1 \over 32 e^2 } Tr [ U^+\partial_\mu U,
U^+\partial_\nu U ] ^2 \],
\end{eqnarray}
where $f_\pi$ is the pion decay constant, $e$ is a dimensionless parameter and $U$ is a SU(2) matrix. 
Performing the collective semi-classical 
expansion\cite{ANW} just substituting $U(r,t)$ by $U(r,t)=A(t)U_0(r)A^+(t)$ in Eq. (\ref{Sky}), being $A$ a SU(2) matrix, we obtain 

\begin{equation}
\label{Lag}
L = - M + \lambda \,Tr [ \partial_0 A\partial_0 A^{-1} ],
\end{equation}
where $M$ is the soliton mass and $\lambda$ is the moment of inertia\cite{ANW}. The SU(2) matrix $A$ can be written as $A=a_0 +i a\cdot \tau$, where $\tau_i$ are the Pauli matrices, and satisfies the spherical constraint relation

\begin{equation}
\label{pri}
T_1 = a_i a_i - 1 \approx 0, \,\,\,\, i=0,1,2,3.
\end{equation}
The Lagrangian (\ref{Lag}) can be read as a function of $a_i$ as

\begin{equation}
\label{cca}
L = -M + 2\lambda \dot{a}_i\dot{a}_i.
\end{equation}
Calculating the canonical momenta

\begin{equation}
\label{cm}
\pi_i = {\partial L \over \partial \dot{a}_i} = 4 \lambda \dot{a}_i,
\end{equation}
and using the Legendre transformation, the canonical Hamiltonian is computed as

\begin{eqnarray}
\label{chr}
H_c=\pi_i \dot a_i-L &=&
 M+2 \lambda \dot a_i\dot a_i \nonumber \\
&=&M+{1\over 8 \lambda } \sum_{i=0}^3 \pi_i\pi_i.
\end{eqnarray}
From the temporal stability condition of the spherical constraint, Eq.(\ref{pri}), we get the secondary constraint

\begin{equation}
\label{T2}
T_2 = a_i\pi_i \approx 0 \,\,.
\end{equation}
We observe that no further constraints are generated via this iterative procedure. $T_1$ and $T_2$ are the second class constraints with

\begin{equation}
\label{Pa}
\{T_1,T_2\}= 2 a_i a_i.
\end{equation}

Using the BFFT formalism we obtain the first class constraints written as\cite{1997} 
\begin{eqnarray}
\label{f1}
\tilde{T}_1&=&T_1+b_1=a_ia_i-1+b_1,\\
\label{f2}
\tilde{T}_2&=&T_2-a_ia_i b_2=a_i\pi_i-a_ia_i b_2,
\end{eqnarray}
which satisfy an Abelian first class algebra
\begin{equation}
\{\tilde{T}_1,\tilde{T}_2\}=0,
\end{equation}
with the WZ variables obeying the following Poisson bracket relation

\begin{equation}
\{b_i,b_j\}=2\epsilon_{ij},\;i,j=1,2.
\end{equation}
The first class Hamiltonian is given by
\begin{eqnarray}
\label{fch}
\tilde{H}&=&M+{1\over 8\lambda}\,
{a_ia_i\over a_ia_i+b_1}\pi_j\pi_j
-{1\over 4\lambda}\,
{a_ia_i b_2\over a_ia_i+b_1}a_j\pi_j
+ {(a_ia_i)^2 (b_2)^2\over a_ia_i+b_1}\nonumber\\\nonumber \\
&=&M+{1\over 8\lambda}\,{a_ia_i\over a_ia_i+b_1}\[\pi_j-b_2 a_j\]^2,
\end{eqnarray}
which also satisfies an Abelian first class algebra
\begin{eqnarray}
\{\tilde{H},\tilde{T}_\alpha\}=0, \;\; \alpha=1,2.
\end{eqnarray}

At this stage, we are ready to apply our formalism. We begin by choosing the first class constraint, Eq.(\ref{f1}), to be the extended gauge symmetry generator

\begin{equation}
\label{esg}
\tilde{T}=\tilde{T}_1=T_1+b_1=a_ia_i-1+b_1.
\end{equation}
The infinitesimal gauge transformations of the WZ variables generated by the extended gauge symmetry generator $\tilde{T}$ are

\begin{eqnarray}
\label{db1}
\delta b_1&=&\epsilon\{b_1,\tilde{T}\}=\epsilon\{b_1,a_ia_i-1+b_1\}=0,\\
\label{db2}
\delta b_2&=&\epsilon\{b_2,\tilde{T}\}=\epsilon\{b_2, a_ia_i-1+b_1\}=-2\,\epsilon.
\end{eqnarray}
From the infinitesimal gauge transformations, Eq.(\ref{db1}), we can choose a representation for $b_1$ as

\begin{equation}
\label{rb1}
b_1=0.
\end{equation}
A representation for $b_2$ can be determined by imposing the first class strong equation, Eq.(\ref{f2})

\begin{eqnarray}
\label{rb2}
a_i\pi_i-a_ia_i b_2=0 \Rightarrow\; b_2={a_i\pi_i\over a_ja_j}.
\end{eqnarray}
As we can see, the function for $b_2$ satisfies the infinitesimal gauge transformation, Eq.(\ref{db2}),

\begin{equation}
\delta b_2=\epsilon\{b_2,\tilde{T}\}=\epsilon\{{a_i\pi_i\over a_ja_j},T_1\}=\epsilon\{{a_i\pi_i\over a_ja_j},a_ia_i-1\}=-2\,\epsilon.
\end{equation}
Then, substituting the functions for $b_1$ and $b_2$, Eqs.(\ref{rb1}) and (\ref{rb2}), in the first class Hamiltonian, Eq.(\ref{fch}), we obtain a gauge invariant Hamiltonian written only in terms of the original second class phase space variables

\begin{equation}
\label{HC}
\tilde{H}=M+{1\over 8\lambda}\[\pi_j\pi_j-{(a_i\pi_i)^2\over a_ja_j}\]=M+{1\over 8\lambda}\[\pi_j\pi_j-{{(T_2)}^2\over a_ja_j}\],
\end{equation}
with the only gauge symmetry generator of the theory 

\begin{equation}
\label{gsgs}
T_1=a_ia_i-1,
\end{equation}
which satisfies an Abelian first class algebra

\begin{eqnarray}
\{ T_1, T_1 \} =0,\\
\{\tilde{H},T_1\}=0.
\end{eqnarray}
Note that when we make the second class constraint equal to zero, $\,T_2=a_i\pi_i=0\,$, we observe that the first class Hamiltonian, Eq.(\ref{HC}), reduces to the original second class Hamiltonian, Eq.(\ref{chr}). Further, the gauge invariant Hamiltonian, Eq(\ref{HC}), is the same obtained when we use two different approaches\cite{GS,GSS}. These results, certainly, indicate the validity of our formalism. 

The gauge invariant Hamiltonian, Eq.(\ref{HC}), can be written as

\begin{equation}
\label{HCM}
\tilde{H}=M+{1\over 8\lambda} \pi_i M^{ij} \pi_j,
\end{equation}
where the phase space metric  $\,M^{ij}\, $ given by

\begin{equation}
\label{M}
M^{ij}=\delta^{ij} - {a^ia^j\over a_ia_i},
\end{equation}
is a singular matrix which has $\,a_i\,$ as an eigenvector with null eigenvalue, namely,

\begin{equation}
a_i M^{ij}=0.
\end{equation}
Then, due to the fact that the matrix $\,M\,$ is singular, in principle, it is not possible to obtain the first class Skyrmion Lagrangian written only in terms of the original second phase space variables with the gauge symmetry generator being $\,T_1\,$, Eq.(\ref{gsgs}).

Now we choose the other first class constraint, Eq.(\ref{f2}), to be the extended gauge symmetry generator of theory

\begin{equation}
\label{esg2}
\tilde{T}=\tilde{T}_2=T_2-a_ia_ib_2=a_i\pi_i-a_ia_ib_2.
\end{equation}
The infinitesimal gauge transformations of the WZ variables generated by this extended gauge symmetry generator $\tilde{T}$ are

\begin{eqnarray}
\label{db12}
\delta b_1&=&\epsilon\{b_1,\tilde{T}\}=\epsilon\{b_1,a_i\pi_i-a_ia_ib_2\}=-2\epsilon a_ia_i,\\
\label{db22}
\delta b_2&=&\epsilon\{b_2,\tilde{T}\}=\epsilon\{b_2, a_i\pi_i-a_ia_ib_2\}= 0.
\end{eqnarray}
From the infinitesimal gauge transformations, Eq.(\ref{db22}), we can choose a representation for $b_2$ as

\begin{equation}
\label{rb22}
b_2=0.
\end{equation}
A representation for $b_1$ can be obtained by imposing the first class strong equation, Eq.(\ref{f1})

\begin{eqnarray}
\label{rb21}
a_ia_i-1+b_1=0 \Rightarrow\; b_1=1-a_ia_i.
\end{eqnarray}
The function for $b_1$ satisfies the infinitesimal gauge transformation, Eq.(\ref{db12}),

\begin{equation}
\delta b_1=\epsilon\{b_1,\tilde{T}\}=\epsilon\{1-a_ia_i,a_i\pi_i\}=-2\,\epsilon.
\end{equation}
Substituting the functions for $b_1$ and $b_2$, Eqs.(\ref{rb22}) and (\ref{rb21}), in the first class Hamiltonian, Eq.(\ref{fch}), we get a gauge invariant Hamiltonian written only in terms of the original second class phase space variables

\begin{equation}
\label{HC2}
\tilde{H}=M+{1\over 8\lambda} \,a_ia_i\pi_j\pi_j,
\end{equation}
with the only gauge symmetry generator of the theory 

\begin{equation}
\label{t2}
T_2=a_i\pi_i,
\end{equation}
which satisfies an Abelian first class algebra

\begin{eqnarray}
\{ T_2, T_2 \} =0,\\
\{\tilde{H},T_2\}=0.
\end{eqnarray}
Again, when we make the second class constraint equal to zero, $\,T_1=a_ia_i-1=0\,$, the gauge invariant Hamiltonian Eq.(\ref{HC2}) reduces to the original second class Hamiltonian, Eq.(\ref{chr}).

The first class Skyrmion Lagrangian can be deduced by performing the inverse Legendre transformation

\begin{equation}
\label{LT}
L=\pi_i\dot{a}_i-\tilde{H},
\end{equation}
where the momentum $\pi_i$ is eliminated by using the Hamilton equation of motion

\begin{equation}
\label{HEM}
\dot{a}_i=\{a_i,\tilde{H}\}={1\over 4\lambda} a_ja_j\pi_i.
\end{equation}
Using relation (\ref{HEM}) in Eq.(\ref{LT}) we derive the first class Lagrangian written as

\begin{equation}
\label{fcl2}
L=-M+ 2 \lambda {\dot{a}_i\dot{a}_i\over a_ja_j},
\end{equation}
with the infinitesimal gauge variation given by $\,\delta a_i=\epsilon\, a_i$, where $\,\epsilon$ is a constant. Notice that it is only possible to derive this first class Lagrangian, Eq.({\ref{fcl2}), if we adopt the symmetry generator of the theory as $\,T_2=a_i\pi_i$, Eq.(\ref{t2}). 

Along the text we have mentioned an important property in our theory which we have only one gauge symmetry generator. Thus from this property we can obtain a second class system from the gauge condition

\begin{equation}
\label{t1theta}
T_{1\theta}=a_ia_i-\theta \, \pi_i\pi_i-1,
\end{equation}
where $\theta$ is a constant. $T_{1\theta}$ is a deformed spherical constraint with the Poisson bracket

\begin{equation}
\{ T_2,T_{1\theta}\}=-2a_ia_i-2\theta \pi_i\pi_i.
\end{equation}
It is not difficult to observe that no additional constraints are generated by imposing the deformed spherical condition relation (\ref{t1theta}). $T_2$ and $T_{1\theta}$ are now the total second class constraints of the model. Using the Dirac brackets formula \cite{rep,Dirac} 

\begin{equation}
\{A,B\}_{DB}=\{A,B\}+{1\over \{T_2,T_{1\theta}\}}\(\{A,T_2\}\{T_{1\theta},B\}-\{A,T_{1\theta}\}\{T_2,B\}\),
\end{equation}
we obtain the commutation relations between the collective coordinates operators upon quantization

\begin{eqnarray}
\label{nca}
\{a_i,a_j\}_{DB}&=&\theta\, {a_j\pi_i-a_i\pi_j\over a^2+\theta\, \pi^2},\\
\{a_i,\pi_j\}_{DB}&=&\delta_{ij}-{a_ia_j+\theta\, \pi_i\pi_j\over a^2+\theta \,\pi^2}, \\
\{\pi_i,\pi_j\}_{DB}&=& {a_j\pi_i-a_i\pi_j\over a^2+\theta\, \pi^2}.
\end{eqnarray}
Note that if we make $\theta=0$ we recover the usual algebra of this collective coordinates operators\cite{JW2}. It is important to observe that using the Dirac bracket quantization we get a noncommuting collective coordinates operators, relation(\ref{nca}). This new result is only derived if we have used $\,T_2=a_i\pi_i$ as the symmetry generator in the first class Skyrmion system.

\section{Conclusions}

In this Letter, we give some prescriptions in order to eliminate the WZ fields of the BFFT formalism. The WZ variables are considered only as auxiliary tools that enforce symmetries in an initial second class constrained system. Then, after embedding a second class system by the BFFT formalism, we substitute the WZ fields by convenient functions and, consequently, we derive a gauge invariant Hamiltonian written only in terms of the original second class phase space variables. This first class system has one gauge symmetry generator. It is an advantage because we have the possibility to select one gauge condition which can reveal important physical properties. 
In the same manner of the BFFT formalism, there is arbitrariness in our prescription. For example, the choices of the extended gauge symmetry generator (and, consequently, the gauge symmetry generator of the theory) and the key function 
$F_\alpha(q_i,p_i)$, Eq.(\ref{fct}), are arbitrary. These different possible options can be used in order to unveil important physical results or some choices can be related to obtain benefits in the algebraic calculations.

\section{ Acknowledgments}
The author would like to thank A.G. Sim\~ao for critical reading. This work is supported in part by FAPEMIG, Brazilian Research Agency.

\end{document}